\documentclass[a4paper]{jpconf}
\usepackage{graphicx}

\newcommand\ket[1]{\left|#1\right>}
\newcommand{\rb}{\mathbf{r}}
\newcommand{\sx}{\mathbf{s}}
\newcommand{\xb}{\mathbf{x}}
\newcommand{\qb}{\mathbf{q}}
\newcommand{\Dx}{\Delta\mathrm{x}}
\newcommand{\nucl}{\mathrm{nucl}}
\newcommand{\eq}{\mathrm{eq}}

\begin{document}
\title{Gravity-related wave function collapse: mass density resolution}

\author{Lajos Di\'osi}

\address{Wigner Research Center for Physics, H-1525 Budapest 114, P.O.Box 49, Hungary}

\ead{diosi.lajos@wigner.mta.hu}

\begin{abstract}
\end{abstract}
Selected issues of the concept of spontaneous collapse are discussed, with the
emphasis on the gravity-related model. We point out that without spontaneous
collapses the Schr\"odinger cat states would macroscopically violate the
standard conservation laws even in the presence of environmental noise.
We prove that the collapse time of condensed matter c.o.m. superpositions is not
sensitive to the natural uncertainty of the nuclear locations whereas
we formulate the conjecture that superfluid He may show an anomalous low
rate of spontaneous collapse compared to common condensed matter.

\section{Introduction}
\label{S1}
If we apply quantum mechanics for macroscopic variables, e.g., to the c.o.m. 
motion of a given massive object, we face paradoxical as well as concrete 
physical issues. The hypothesis of spontaneous collapse (SC) has been thought a remedy. 
According to it, all macroscopic superpositions, i.e.: the Scr\"odinger cat (Cat) states, 
are quickly eliminated as if a quantum measurement were performed on them. This time,
however, the measurement happens spontaneously and universally, without the 
presence of any measurement device. SC models contain
a further innovation compared to standard wave function collapse. The
latter is a sudden one-shot event whereas SCs take time,
they have a certain stochastic dynamics. Formally, they are
time-resolved standard one-shot collapses. The SC
models take out the concept of standard one-shot collapse.     

We discuss some old problems within the gravity-related model, 
proposed independently by Penrose and the author \cite{Dio86,Dio87,Dio89,Pen94,Pen96,Pen98,Pen04}, 
of the yet hypothetical SC. Some problems are acute because of related experimental 
considerations and experiments \cite{Mar_Bou03,Adl07,Rom11,LiKheRai11,Van_Asp11,Pep_Bou12,Yan_etal12}.
Some are common for almost all SC models \cite{BasGhi03}, some are specific for ours. 
   For the sake of a minimum comparison, we introduce the elements of the gravity-related SC
model together with the so-called continuous spontaneous localization (CSL) \cite{mCSL} model, 
using the notion of `catness' (Sec.~\ref{S2}).    
   Sec.~\ref{S3} presents two novel results for the gravity-related model.
First, we show that the equilibrium collapse rate of the Cat c.o.m. is
a classical expression, independent of $\hbar$. Second, we prove that
the calculated SC rate does not depend on the spread
of nuclear locations even if the mass density spatial resolution assumes
the nuclear size.
 Many question the physical relevance of SCs of
the Cat states because their unavoidable entanglement with the
environment prevents Cat states from being created. We point out in
Sec.~\ref{S4}, that the immediate environmental decoherence may not help
Cat states avoid their macroscopic violation of conservation laws.

\section{Spontaneous Collapse Models}
\label{S2}
Most SC models are similar to each other. Let us introduce
the notion of `catness' for the superposition of two different 
macroscopic configurations, say, two mass densities $f(\rb)$ and $f'(\rb)$.
Consider a typical Cat state:
\begin{equation}
\ket{\mathrm{Cat}}=\ket{f}+\ket{f'},
\label{Cat_f}
\end{equation}
where $f$ and $f'$ are `macroscopically' different.
Macroscopicity, i.e.: catness, is measured by a distance $\ell(f,f')$.
Let, for concreteness, $\ell^2$ be of dimension of energy. 
SC models assume that Nature makes $\ket{\mathrm{Cat}}$ decay 
(collapse) stochastically at mean life-time
\begin{equation}
\tau=\frac{\hbar}{\ell^2(f,f')}.
\label{tau}
\end{equation}
Each collapse model has its own choice of the measure
$\ell$ of catness. The choice of the distance $\ell(f,f')$ must be careful:
there should be practically no decay (extreme large life-time $\tau$)  
for atomic cats, and immediate decay (very short life-time $\tau$) 
for `macroscopic' Cats.
When specifying catness, the spatial mass density resolution is a crucial issue. 
The models set up a finite cutoff $\sigma$ of spatial resolution. 

The gravity-related (G-related, or DP) model defines
\begin{equation}
\ell_G^2(f,f')=G\int\int [f(\rb)-f'(\rb)][f(\sx)-f'(\sx)]\frac{d\rb d\sx}{\vert\rb-\sx\vert},
\label{ell2_G}
\end{equation}
where $f,f'$ are coarse-grained with the spatial resolution $\sigma$. The model is further
discussed in Sec.~\ref{S3}.
The closest alternative to the DP model is the mass-proportional 
CSL model \cite{mCSL}. The CSL choice of catness reads
\begin{equation}
\ell^2_{CSL}(f,f')=~\frac{\hbar\lambda\sigma^3}{m_0^2}\int [f(\rb)-f'(\rb)]^2 d\rb,
\label{ell2_CSL}
\end{equation}
where $f,f'$ are Gaussian coarse-grained with the spatial resolution $\sigma\!\!\sim\!\!10^{-5}$cm.
The CSL model uses the nucleon mass $m_0$ as a reference mass and the further 
fenomenological parameter $\lambda\!\!\sim\!\!10^{-17}/$s.

The predictions of the DP and CSL models may differ by orders of magnitudes, but they
fulfill the main request. Practically infinite life-times are preserved 
for superpositions of microscopic systems whereas the Cat states
(those apparently missing from Nature) will decay immediately. According to both
DP and CSL models, the competition of SCs with the standard unitary
dynamics may reach a regime of equilibrium. The DP model is unique in that the
equilibrium collapse rate is classical, $\hbar$ cancels from it (Sec.~\ref{S3.1}).

\section{G-related spontaneous collapse}
\label{S3}
\begin{figure}[h]
\begin{center}
\includegraphics[width=.3\textwidth]{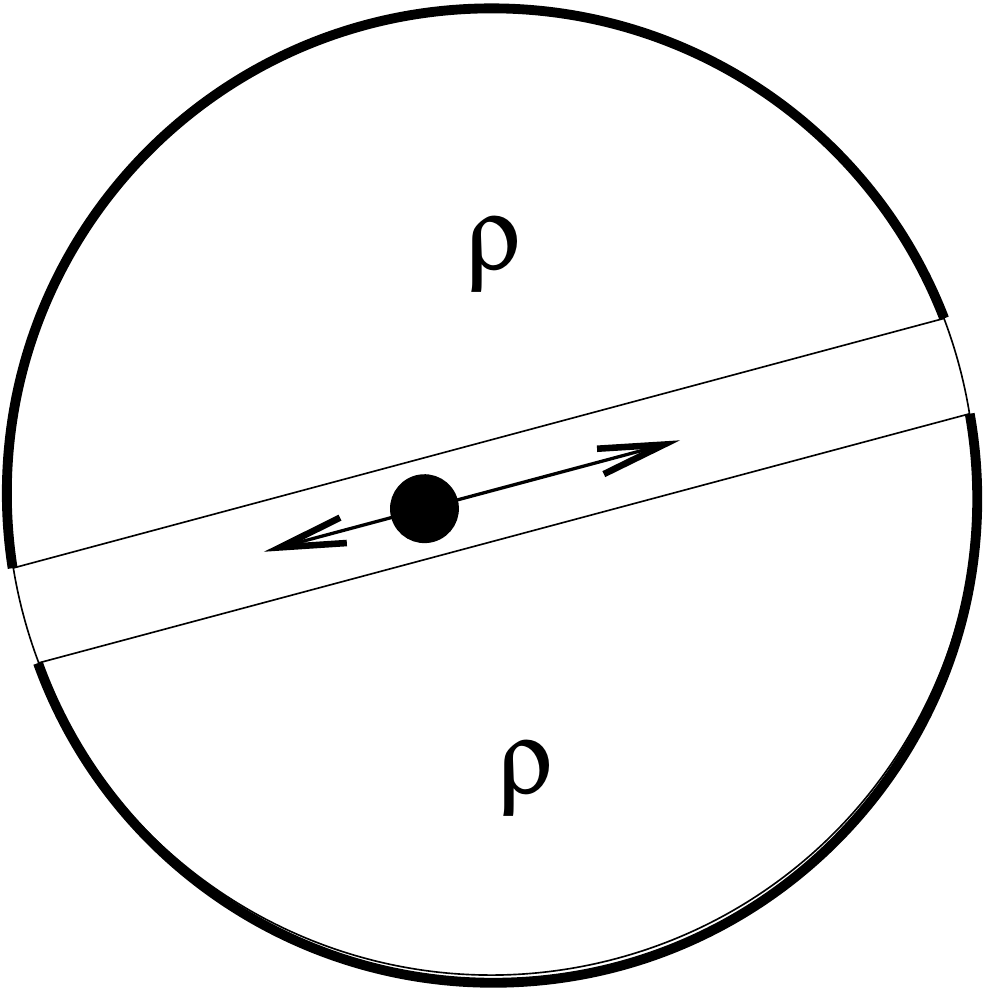}\hspace{2pc}
\begin{minipage}[b]{14pc}\caption{\label{Fig}Schematic view of the Newton Oscillator: a homogeneous ball with an infinite
narrow diagonal hole where the probe is oscillating under the linear directional force of the Newton 
field of the ball. The frequency of the oscillator $\omega_G=\sqrt{4\pi G\rho/3}$ does not depend on
the size but on the density $\rho$ of the ball.}
\end{minipage}
\end{center}
\end{figure}

We can re-express the (squared) catness (\ref{ell2_G}) as the following combination
of three Newton interaction energies:  
\begin{equation}
\ell_G^2(f,f')=2U(f,f')-U(f,f)-U(f',f'),
\label{ell2_G_U}
\end{equation}
where 
\begin{equation}
U(f,f')=-G\int\int\frac{f(\rb)f'(\sx)}{\vert\rb-\sx\vert}d\rb d\sx.
\label{Uff}
\end{equation}
Suppose our system is a single rigid homogeneous ball of mass $M$ and radius $R$
and we consider its c.o.m. wave function. The popular 'mechanical' Cat 
state \cite{Dio86,Dio87,Dio89,Pen94,Pen96,Pen98,Pen04,Adl07} is the superposition of 
two macroscopically different c.o.m. locations $\xb$ and $\xb'$:
\begin{equation}
\ket{\mathrm{Cat}}=\ket{\xb}+\ket{\xb'}.
\label{Cat}
\end{equation}
The corresponding two mass densities of the Cat state (\ref{Cat_f}) read
\begin{eqnarray}                                                      
f(\rb)&=&\rho\theta(\vert\rb-\xb\vert\leq R),\nonumber\\              
f'(\rb)&=&\rho\theta(\vert\rb-\xb'\vert\leq R),                       
\label{f_R}                                                           
\end{eqnarray}                                                        
where $\rho=M/(4\pi R^3/3)$ is the constant mass density, $\theta$ is the step-function.
The central quantity                                                                    
is the mean collapse time $\tau_G$, cf. (\ref{tau}), of the c.o.m. wave function.                          
It becomes the function of  the c.o.m.                                                  
displacement $\Delta\xb=\xb-\xb'$; we calculate the mean collapse rate in the first non-vanishing order \cite{Dio86,Dio87,Dio89}:    
\begin{equation}                                                                        
\frac{1}{\tau_G}=\frac{1}{2}\frac{M\omega_G^2}{\hbar}(\Delta\xb)^2,                    
\label{rate}
\end{equation}
valid if $\vert\Delta\xb\vert\ll R$.
The $R$-dependence has been absorbed into the parameter
\begin{equation}
\omega_G=\sqrt{4\pi G\rho/3}
\label{om_G}
\end{equation}
which we call the frequency of the Newton oscillator (cf. Fig.~\ref{Fig}).
This frequency is purely classical. 

\subsection{Equilibrium collapse rate}
\label{S3.1}
The existence of equilibrium is a generic feature of the SC models.
We are going to use the heuristic arguments of \cite{Dio87} which 
anticipated the exact derivations \cite{Dio89}.

Let us start from the collapse rate (\ref{rate}) of a mechanical 
Cat, previously determined in the leading order of the coherent displacement $\Delta\xb=\xb-\xb'$.
We can heuristically extend this expression for a c.o.m. wave packet of coherent
dispersion $\Dx$: the SCs at about the rate (\ref{rate}) tend to
shrink the width $\Dx$. Now we take the unitary evolution of the wave packet
into the account: it tends to spread $\Dx$ at the rate about $\hbar/M(\Dx)^2$.
The two contrary tendencies get balanced when the two rates coincide:
\begin{equation}
\frac{\hbar}{M(\Dx)^2}\sim\frac{M\omega_G^2(\Dx)^2}{\hbar}.
\label{balance}
\end{equation}
This is the condition of equilibrium between SC and unitary dynamics,
resulting in a sort of natural quantum state of the massive object. 
The corresponding natural coherent size of the c.o.m. wave packet yields
\begin{equation}
\Dx^\eq \sim \sqrt{\hbar/M \omega_G},
\label{Dx_eq}
\end{equation}
which is a very tiny size for massive objects due to the smallness of $\hbar$.
Similar strong localization of the macro-objects follows from the CSL model, too. 
Nevertheless, we find a unique feature of the DP model if we calculate
the equilibrium collapse rate. Let's just take the geometric mean of both
sides of the balance condition above, it yields the equilibrium collapse rate
directly: 
\begin{equation}
\frac{1}{\tau_G^\eq}\sim\omega_G.
\label{rate_eq}
\end{equation}
The equilibrium collapse rate coincides approximately with the frequency of the Newton oscillator,
does not depend on the mass $M$ and size $R$, moreover, it is fully classical, independent of $\hbar$. 

\subsection{Mass resolution issue, He superfluid}
\label{S3.2}
Catness (\ref{ell2_G}) and collapse rates would diverge for point-like objects, that's why
we need the finite spatial resolution $\sigma$ of $f(\rb)$. It's crucial to decide that we resolve
the microscopic structure or we don't. 
Adopting the resolution $\sigma\!\!\sim\!\!10^{-5}$cm of the
CSL model, the mass density in condensed matter becomes the smooth macroscopic density 
$\rho$ equal to a few times $1$g/cm$^3$. We can estimate the parameter $\omega_G$ which
governs the collapse rate (\ref{rate}) and yields the equilibrium collapse rate (\ref{rate_eq}):
 \begin{equation}
\frac{1}{\tau_G^\eq}=\omega_G^\eq\sim\frac{1}{\mathrm{h}}.
\end{equation}
This extreme low rate questions the significance of DP collapses in common objects.  
The microscopic structure of the mass density may be a loophole. 

We can take a much finer resolution with the 'nuclear' cutoff $\sigma\!\!\sim\!\!10^{-12}$cm.
Then $f(\rb)$ shows the granular structure of localized nuclear contributions. 
For simplicity, we consider that the mass $M$ is carried by the nuclei which are homogeneous 
balls of radius $\sigma$ and density $\rho^\nucl=(R/\sigma)^3\rho$. For small
displacements $\vert\Delta\xb\vert\!\!\ll\!\!\sigma$, each nucleus will contribute separately to
the (squared) catness, like a homogeneous ball of radius $\sigma$ and density $\rho^\nucl$.
At the end of the day we obtain the collapse rate similar to (\ref{rate}):
\begin{equation}
\frac{1}{\tau_G}=\frac{1}{2}\times\frac{M(\omega_G^\nucl)^2}{\hbar}(\Delta\xb)^2,
\label{rate_micro}
\end{equation}
just with $\omega_G^\nucl$ instead of $\omega_G$ \cite{Dio07}. 
The expression is valid if $\vert\Delta\xb\vert\ll\sigma$.
Here
\begin{equation}
\omega_G^\nucl=\sqrt{4\pi G\rho^\nucl/3}
\label{om_G_nucl}
\end{equation}
is the frequency of the Newton oscillator in nuclear matter.
Its order of magnitude is $\omega_G^\nucl\!\!\sim\!\!10^3/s$.

With the 'nuclear' resolution, the Cat life-time $\tau_G$ becomes cca $10^{12}$ times shorter! 
Without this amplification, first shown in \cite{Dio07}, the experimental significance of 
the SCs may remain illusory for another while, as recognized in \cite{Rom11,LiKheRai11,Pep_Bou12,Yan_etal12}.
It seems therefore a sort of necessity that we define the DP model with the 'nuclear' cutoff to
take granular structure of the mass density $f(\rb)$ into the account. 

We must note that the exact interpretation of the nuclear mass distribution does not exist. 
We don't know the principles either. In the best case we believe that high energy gravitons would copy 
how high energy photons explore the electronic charge density. But the available non-relativistic
interpretation of nuclear mass density is fenomenological, we have to rely upon the simple fenomenology.   

Finally, we ought to discuss whether or not 
the thermal and quantum blurredness of the nuclear locations matters when we calculate the DP collapse
rate of the Cat state. In solids, the blurredness of nuclear location is about one tenth of the interatomic 
distance. If this broadening of mass distribution did matter then the above said factor $10^{12}$ of 
amplification would be lost and replaced by some $10^3$. Fortunately, this is not the case: 
the DP collapse of the c.o.m. superposition is always independent of the quantum spread of 
microscopic constituents. The general proof is trivial in the density matrix formalism. 
Let us introduce the coordinates $\qb_1,\qb_2,\dots,\qb_N$ of the $N$ nuclei w.r.t. the c.o.m. $\xb$. 
We choose the following set of independent coordinates: $\xb,\qb_1,\qb_2,\dots,\qb_{N-1}$.
The corresponding density matrix reads
\begin{equation}
\rho(\xb,\qb_1,\qb_2,\dots,\qb_{N-1}\vert\xb',\qb_1',\qb_2',\dots,\qb_{N-1}').
\end{equation}
The full DP collapse rate of this state of the Cat is described by the following term \cite{Dio86,Dio87,Dio89}:
\begin{equation}
\frac{\ell_G^2(f,f')}{\hbar}\rho(\xb,\qb_1,\qb_2,\dots,\qb_{N-1}\vert\xb',\qb_1',\qb_2',\dots,\qb_{N-1}'),
\label{rate_tot}
\end{equation}
where $f$ is the ($\sigma$-coarse-grained) mass density of the configuration $(\xb,\qb_1,\qb_2,\dots,\qb_{N-1})$ and 
$f'$ is the ($\sigma$-coarse-grained) mass density of the configuration $(\xb',\qb_1',\qb_2',\dots,\qb_{N-1}')$. 
We are not interested in the collapse rate of the integral degrees of freedom, therefore we trace over them, yielding  
\begin{equation}
\int\frac{\ell_G^2(f,f')}{\hbar}\rho(\xb,\qb_1,\qb_2,\dots,\qb_{N-1}\vert\xb',\qb_1,\qb_2,\dots,\qb_{N-1})
      d\qb_1 d\qb_2\dots d\qb_{N-1},
\label{rate_com}
\end{equation}
where $f$ is the ($\sigma$-coarse-grained) mass density of the configuration $(\xb,\qb_1,\qb_2,\dots,\qb_{N-1})$ and 
$f'$ is the ($\sigma$-coarse-grained) mass density of the configuration $(\xb',\qb_1,\qb_2,\dots,\qb_{N-1})$. 
The expression (\ref{rate_com}) is the exact c.o.m. collapse rate.
This shows that the coherence of the spread of the nuclear locations is completely irrelevant.
And there is an even more important result. We must not blur the nuclear locations first.
Rather we must calculate the c.o.m. squared catness and the corresponding collapse rate 
with the sharp locations $\qb_1,\qb_2,\dots$ of the nuclei. 
Second,  the effect of the spread of nuclear locations will be taken into the account 
by blurring the above squared catness (i.e., the collapse rate) over the spread of the nuclear locations.
In our special case of very small displacements $D\xb\ll\sigma$, the squared catness 
$\ell_G^2(f,f')$ is, as we said before, well approximated by the sum of separate contributions of the nuclei. 
The contribution of a single nucleus, say the 'first' one, depends on the displacement between the two superposed locations 
$\xb+\qb_1$ and $\xb'+\qb_1$, which is just $\vert\Delta\xb\vert=\vert\xb-\xb'\vert$.    
Neither the precise nuclear locations nor their spreads will influence the c.o.m. SC of the Cat.       

Liquid He is an exception. In superfluid He the quantum dispersion of the nuclei is of the order of
the interatomic distance, the nuclei have no identity and the above proof does not apply.
Distinction between internal and c.o.m. collapses, respectively, is more involved.
In liquid He the c.o.m. collapse rate may be much smaller than the value
(\ref{rate_micro}) valid in common condensed matter. Superfluid He may have an anomalous
low rate of SCs in it. Of particular interest is the temperature dependence
of the collapse rate. Too high temperatures would push back the coherent dispersion
of the nuclei below the interatomic scale, resulting in increased c.o.m. collapse rate. 
It is a further question yet to be answered whether cold liquid He is sufficient for the anomalous 
low collapse rates or superfluidity is essential for it.     

\section{Environment does not mask spontaneous collapse}
\label{S4}
We know that typical environmental noise and entanglement used to decohere any 
Cat state much earlier than SCs would kill it.
Environmental noise in Nature used to be much stronger than SCs.
The experimental test of SCs appears very hard.
Since SCs are masked by environmental noise, one might think
that SCs are irrelevant, the corresponding models are superfluous.
This conclusion is not correct.

Let us concentrate on the mechanical Cat. 
In the naive Cat state (\ref{Cat}) we considered the c.o.m. 
but ignored all internal and environmental degrees of freedom.
Now we refine the naive Cat state and include the environmental degrees of freedom:
\begin{equation}
\ket{\mathrm{Cat}}=\ket{\xb}\otimes\ket{E}+\ket{\xb'}\otimes\ket{E'},
\label{Cat_E}
\end{equation} 
where $\ket{E}$ and $\ket{E'}$ stand for two different states of the environment. 
To distill the central issue, we assume that the only massive object is the Cat
itself (a rigid homogeneous ball with c.o.m. $\xb,\xb'$) and the environment
is massless compared to the Cat. The above state is a macroscopic superposition 
of different mass configurations (\ref{f_R}), so it is a genuine Cat. 
The two states $\ket{\xb}$ and $\ket{\xb'}$ 
are entangled with the environmental states $\ket{E}$ and $\ket{E'}$, respectively. 
Since in practice the environmental states $\ket{E}$ and $\ket{E'}$ are orthogonal, 
the coherence of the macroscopic superposition is not observable locally, 
our Cat is not observable locally. To see the Cat, we need to observe the remote 
parts of the entangled environment that we are not able to do. 
How can the Cats keep annoying us? Why should we deploy SCs or
any mechanism against them at all?

Suppose the Cat state in the following form:
\begin{equation}
\ket{\mathrm{Cat}}=\ket{\xb}\otimes\ket{E}\otimes\ket{1}+\ket{\xb'}\otimes\ket{E'}\otimes\ket{2},
\label{Cat_E12}
\end{equation} 
where $\ket{1},\ket{2}$ are orthogonal single photon states. Say, the photon
(like billions of other environmental photons) bounced back from the Cat some time ago and went to the states 
$\ket{1},\ket{2}$ entangled with the Cat c.o.m. states $\ket{\xb},\ket{\xb'}$, respectively.
Suppose that we measure the photon so that, according to the standard quantum measurement
theory, the Cat collapses into $\ket{\xb}$ or into $\ket{\xb'}$ at random.
The c.o.m. of the Cat (and of the whole system) 
is $(\xb+\xb')/2$ before the measurement. After the measurement it becomes either
$\xb$ or $\xb'$. We got a macroscopic shift of the c.o.m. of a closed macroscopic system
whose only interaction with the outer world was that we measured a single photon that
belonged to the system. This is not merely paradoxical. Actually all (non-superselected) 
conservation laws are macroscopically violated by the mere existence of certain Cats whether 
they are naive isolated ones or real ones entangled with the environment. Violation of 
energy-momentum conservation in standard collapse has been discussed earlier \cite{Pea00}. Note that
our present concern is the possible macroscopicity of the violation, not the violation itself. 
A microscopic defect of conservation can always be attributed to (and balanced by) the 
measuring device. But macroscopic defects are annoying.

The SC models, like DP and CSL, offer a simple treatment of the
Cat problem. They collapse the Cats (\ref{Cat_E}), kill them before they can come up, 
much before we could measure them and provoque the macroscopic violation of conservation laws. 
The collapse models themselves impose violation of the conservation laws at a tinier scale, 
but they prevent macroscopic violations at least.    

\section{Closing remarks}
\label{S5}
We have added some new facts, thoughts and conjectures to the gravity-related SC model. 
On one end, sound analysis has clarified how the thermal and quantum motion of the nuclei 
constituting a massive body should enter the calculation of the c.o.m. collapse rate.  
On the other extreme, we have conjectured that in superfluid He the c.o.m. 
SC rate may be much lower than in common condensed matter. 
Ref.~\cite{Dio13} by the present author discusses similar topics and a related possible 
extension of the DP model.  

\ack
Support by the Hungarian Scientific Research Fund under Grant No. 75129
and support by the EU COST Action MP100 are acknowledged.

\end{document}